\documentclass[pra,aps,twocolumn,showpacs,superscriptaddress,amsmath,amssymb]{revtex4}
\usepackage{color}
\usepackage{amsmath,graphicx,bm}

\begin{document}
\def\tr{\rm{Tr}}
\def\la{{\langle}}
\def\ra{{\rangle}}
\def\a{{\alpha}}
\def\e{\varepsilon}
\def\q{\quad}
\def\w{\widetilde{W}}
\def\t{\widetilde{t}}
\def\a{\hat{A}}
\def\h{\hat{H}}
\def\E{\mathcal{E}}
\def\p{\hat{P}}
\def\u{\hat{U}}
\def\ros{R_{\mathrm{so}}}
\def\et{{\bm\eta}}
\def\Ps{{\bm\Psi}}
\def\n{\nonumber}
\def\eti{{\bm\eta}^{I}}

\title{Spin measurements and control of cold atoms using spin-orbit fields}

\author {D. Sokolovski$^{1,2}$ and E. Ya. Sherman}
\affiliation{Departmento de Qu\'imica-F\'isica, Universidad del Pa\' is Vasco, UPV/EHU, Leioa, Spain}
\affiliation{IKERBASQUE Basque Foundation for Science, Bilbao, Spain}

\date{\today}
\begin{abstract}
{We show that by switching} on a spin-orbit interaction {in a cold-atom system, experiencing a Zeeman-like coupling to an external
field, e.g., in a Bose-Einstein condensate,} one can simulate a quantum measurement 
on a precessing spin. 
Depending on the realization,  the measurement can access both the ergodic and the Zeno regimes, 
while {the time dependence} of the spin's decoherence may vary from a Gaussian to an inverse fractional power law. 
{Back action of the measurement forms time- and coordinate-dependent profiles of the atoms'
     density, resulting in its translation, spin-dependent fragmentation, and appearance of 
     interference patterns.} 
\end{abstract}

%
%
\maketitle
\section{Introduction}

Recent advances in producing synthetic spin-orbit coupling fields in cold
atoms, both bosonic \cite{Stanescu08,Lin11} and fermionic, \cite{Liu09,Wang,Cheuk} opened
a new field of research in cold matter physics reviewed in Refs.[\onlinecite{Dalibard,Zhai,Goldman}]. 
Spin-orbit coupling of the Rashba and Dresselhaus
forms, as well as effective magnetic fields leading to the Zeeman-like splitting  
for the corresponding pseudospin, can be designed there by optical means. 
In these systems the pseudospin is formed by coupling hyperfine atomic levels
with highly coherent resonant laser radiation. Since this coupling 
strongly depends on the detuning of the laser frequency from the resonance, the movement of an atom 
also modifies its interaction with light via the Doppler shift
linear in the atom velocity. This effect is seen as the effective spin-orbit coupling.

There are at least two advantages of the synthetic spin-orbit couplings over those observed
in solids \cite{SOLID}. First, the effect of the coupling on cold atoms  can be considerably stronger than that on electrons 
in semiconductors. Second, it may be designed for a particular purpose,  and can be switched on and off as required.
The relative strength of the coupling {can make possible}
observation of new macroscopically ordered phases (see e.g. \cite{PHASE1,PHASE2,PHASE3}),
topological states (e.g. \cite{topo}), and nontrivial spin dynamics \cite{Tokatly,Natu,MWWu}.  
Flexibility in designing the fields suggests new applications of cold atoms
in dynamical systems, for example, developing of new quantum measurement techniques. 
Below we use this coupling in order to simulate a measurement on 
spin precessing in an external {magnetic field},
a fundamental problem in many branches of physics.
Conversely,  spin-orbit coupling can be seen as a means of controlling the motion of atoms.
In particular, we will explore the measurement's ability to
manipulate coherent motion of atoms in a Bose-Einstein
condensate with spin-orbit coupling, where the {effect of the coupling}
is enhanced by a large number of participating bosons in the same quantum
state. 

{The rest of the paper is organized as follows. In Sections II and III we introduce coupled spin-orbit dynamics and
show that it can serve as a quantum meter for the (pseudo)spin degrees of freedom, with the center of mass of the atom 
playing the role of a von Neumann pointer. In Sec. IV we study decoherence produced on the spin by the measurement.
In Sec. V we use simple results from the measurement theory in order to describe the spacial distribution of the atomic cloud
and identify Zeno, ergodic, and intermediate regimes of the measurement. In Section VI we will suggest
possible experimental implementations. Brief conclusions are given in Sec. VII.}

\section{Spin-orbit dynamics}

We begin with the Rashba Hamiltonian for an atom of mass $M$ moving {in one dimension} \cite{BOX,Sinha} in the presence of an 
external magnetic field $\mathbf{B}=B\mathbf{n}$
($\hbar\equiv 1$): 
\begin{equation}\label{1}
\h= \hat{p}^2/2M + \omega (\mathbf{n}\cdot \bm{\sigma})+v_{\rm so}\hat{p}\sigma_{z},
\end{equation}
where $\hat{p}$ is the momentum, {$2\omega$ is the Larmor frequency}, $v_{\rm so}$ is the spin-orbit velocity,
and $\sigma_{i}$ are the Pauli matrices. {The spin-orbit coupling term in Eq.\eqref{1} results  in
breaking the Galilean invariance \cite{GALILEO}, so that 
the system can have a non-zero velocity even for a zero total momentum.} 

The system has a characteristic length, the distance $x_{\rm so}= v_{\rm so}/\omega$ which an atom travels
at $v_{\rm so}$ in $1/\pi$ of the Larmor period.  With the help of $x_{\rm so}$ we define the dimensionless variables 
 $t\to\omega t$, $x\to x/x_{\rm so} $, $M\to Mv_{\rm so}x_{\rm so}$, and $p\to x_{\rm so} p_{}$,
 which we will use below  unless  stated otherwise.
 The wavefunction of an atom with
a pseudo-spin 1/2 is a two-component spinor ${\bm \Psi}(x,t)$ which, in the new variables,
satisfies the Schr\"odinger equation:
\begin{equation}\label{4}
i\partial_t {\bm \Psi}(x,t)=[(\mathbf{n}\cdot \bm{\sigma})-i\partial_x \sigma_z-\partial_x^2/2M] {\bm \Psi}(x,t).
\end{equation}
We assume that the spin-orbit coupling is activated at $t=0$, when the center of mass of the  atom is described by a 
wave packet 
\begin{equation}
G(x,t=0)=\int A(p)\exp(ipx)dp,
\end{equation}
and its spin is in an initial pure state $\et^{[{\rm in}]}$. 
With no loss of generality we consider an atom initially at rest,  so that $|A(p)|=|A(-p)|$ 
is peaked around $p=0$. 
{(An additional constant field due to the atom's nonzero momentum can be absorbed in $\mathbf{B}$.)} 
Finally, we choose $\mathbf{B}$ to lie in the $xz$-plane 
making an angle $\theta$ with the $z$-axis, $\mathbf{n}=(\sin\theta, 0, \cos\theta)$. 

The subsequent dynamics is not trivial: for each momentum $p$, the spin orbit coupling adds an extra 'magnetic' field, 
proportional to $p$, along the $z$-axis,  so that the resultant strength of the field in which the spin precesses, 
is uncertain. The ''anomalous'' spin-dependent contribution to the atom's velocity, $i[\h,x]$, is proportional to the $z$-component of the spin, so 
the atom's translational motion is also complicated. We can write the wave function as an integral over the atom's momenta,
\begin{eqnarray}\label{5}
{\bm \Psi}(x,t)=\int A(p)\exp(ipx-ip^2t/2M) \u(p,t)\et^{[{\rm in}]}dp,\q
\end{eqnarray}
where 
$\u(p,t)\equiv \exp\{-i[(\mathbf{n}\cdot \bm{\sigma})+p\sigma_z]t\}$
is the evolution operator for a spin precessing in the composite field $\widetilde{\mathbf B}$ 
with components $(\sin\theta, 0, \cos\theta + p)$. Its matrix elements are 
\begin{eqnarray}\label{6}
&&U_{11}=U^{*}_{22} = \cos(\Omega t)-\frac{i}{\Omega}(p+ \cos\theta)\sin(\Omega t), 
\\ \nonumber
&&U_{12}=U_{21}=- \frac{i}{\Omega}\sin\theta\sin(\Omega t),
\end{eqnarray}
where
\begin{eqnarray}\label{6a}
\Omega\equiv\Omega(p,\theta)=\sqrt{(p+\cos\theta)^2+\sin^2\theta}.
\end{eqnarray}
Equivalently, ${\bm \Psi}(x,t)$ can be written as a convolution in the coordinate space, 
\begin{eqnarray}\label{7}
{\bm \Psi}(x,t)=\int G(x-x',t)\et(x',t)dx',
\end{eqnarray}
where the wave packet 
\begin{equation}
G(x,t)=\int A(p)\exp(ipx-ip^2t/2M)dp
\end{equation}
describes the motion of the atom without spin-orbit coupling, and the sub-states $\et(x',t)$ are given by
\begin{eqnarray}\label{8}
\et(x,t)=(2\pi)^{-1} \int \exp(ipx)\u(p,t)\et^{[{\rm in}]} dp.
\end{eqnarray}
\newline
Both representations, (\ref{5}) and (\ref{7}) have advantages, which we will explore below.
\section{Spin-orbit system as a quantum meter}
We begin by revisiting classical measurements. Consider a classical system with coordinate $X$ and 
momentum $P$, coupled
 to a pointer, whose coordinate and momentum are $x$ and $p$, 
respectively. For the full Hamiltonian we write
\begin{equation}\label{2}
H(P,X,p,x)=H_0(P,X)+\sigma(P,X)p+p^2/2M,
\end{equation}
where $H_0(P,X)$ is the system's Hamiltonian, and $\sigma(P,X)$ is some dynamical variable of  {interest}  \cite{vN}. 
Setting the pointer to zero, $x(t=0)=0$, and giving it an initial momentum,
$p(t=0)=p_0$, we easily find 
\begin{equation}\label{3}
x(t)/t-p_0/M=t^{-1}\int_0^t \sigma(P(t'),X(t'))dt' \equiv\la \sigma\ra_t ,
\end{equation}
where $\{X(t),P(t)\}$ is the trajectory of  a system governed by the  modified 
Hamiltonian $H_{p_0}=H_0(P,X)+\sigma(P,X)p_0$. Thus, by reading the final pointer position, $x(t)$, 
we also measure the time average $\la \sigma\ra_t$
along a trajectory which is  perturbed by the meter, unless the pointer is at rest, $p_0=0$. 

Equation (\ref{4}) is a quantum version of Eq.(\ref{2}), with the $z$-component of the spin, 
$\sigma_{z}$ playing the role of the measured variable, the center of mass of the atom with the position $x$ playing the 
role of the pointer, and the coupling to the magnetic field - the role of $H_0$. We have, 
therefore, a quantum measurement which we will analyze in detail.
The connection with the classical measurement above becomes even more evident if we  
slice the time interval into $K$ subintervals
of length $\epsilon=t/K$, send $K$ to infinity, and use the Trotter formula to factorize
$\exp\{-i[(\mathbf{n}\cdot \bm{\sigma})+p\sigma_z]\epsilon\}=\exp\{-i(\mathbf{n}\cdot \bm{\sigma})\epsilon\}\exp(-ip\sigma_z\epsilon)$. 
In this way, we obtain for the spin a set of virtual Feynman paths taking the values
$s_k=\pm1$, $k=1,..,K+1$ at each discrete time, and define for each path an 
evolution operator  
\begin{eqnarray}\label{10a}
&&\u(p,t|{\rm path})\equiv  \exp\left[-ip\sum_{k=1}^K s_k\epsilon\right]
\times  \\
&&\hspace{-1cm}|s_{K+1}\ra\la s_{K+1}|\exp[-i(\mathbf{n}\cdot\bm{\sigma})\epsilon]|s_{K}\ra\times \ldots \nonumber \\
&&\hspace{0.5cm}\la s_2|\exp[-i(\mathbf{n}\cdot \bm{\sigma})\epsilon]|s_{1}\ra\la s_1| \nonumber\\
&&=\exp\left[-ip\sum_{k=1}^K s_k\epsilon\right]
\hat{\mathcal{U}}(t,\rm path).
\nonumber  
\end{eqnarray}
The sum in the first exponent is proportional to the time average of $\sigma_z$, defined along a spin's 
Feynman path  in a similar way the time average of $\sigma(P,X)$, $\la \sigma \ra_{t},$  
was earlier defined in Eq.(\ref{3}) for a classical trajectory. 
The resulting time average of $\sigma_z$ along a given path $s_k$ is: 
\begin{eqnarray}\label{9}
\la \sigma_z [{\rm path}]\ra_t=t^{-1}\lim_{K\to \infty} \sum_{k=1}^K s_k\epsilon \equiv t^{-1}\int_0^t s(t')dt'.
\end{eqnarray}
Interchanging integration over $p$ with summation over the paths, we can rewrite Eq.(\ref{8}) as a
sum over only such evolutions for which the time average of $\sigma_z$ equals $x/t$ {in the form}:
\begin{eqnarray}\label{10}
\et(x,t)=
\sum_{\rm paths} 
\delta(t\la \sigma_z\ra_t-x)\hat{\mathcal{U}}(t,\rm path)\et^{[{\rm in}]},
\end{eqnarray}
where $\hat{\mathcal{U}}(t,\rm path)$, acting on the spin's degrees of freedom, is defined in Eq.(\ref{10a}).
This is a quantum analog of Eq.(\ref{3}) - the atom arrives at a location $x$ provided the time average of  $\sigma_z$ is equal to $x/t$.
There are, however, important differences.
First, with many virtual paths involved, $\la \sigma_z\ra_t$ 
is a distributed quantity, whose value may lie between 
$1$ (the spin spends all the time in the $|s=1\ra$ state) and $-1$ (the spin spends all the time in the $|s=-1\ra$ state).
Second, we can only determine $\la \sigma_z\ra_t$ to a finite accuracy.
If $G(x-x',t)$ in Eq.(\ref{7}) is peaked around $x'=x$ with the width $\Delta x$, by
finding the particle in point $x$ one can ascertain the value of $\la \sigma_z\ra_t$ with an error $\sim \Delta x/t$.
Finally, even though the mean momentum of the atom remains zero at all times, a narrow $G(x,0)$ requires a spread of $p$'s 
around $p=0$, and additional fields produced by these non-zero momenta perturb the spin's motion. 
The perturbation is greater for more accurate measurements. It is likely to result in decoherence of the 
initially pure spin state $\et^{[{\rm in}]}$, which will discuss next.
\section{State-dependent decoherence of the spin}

\begin{figure}[h]
\includegraphics*[scale=0.8]{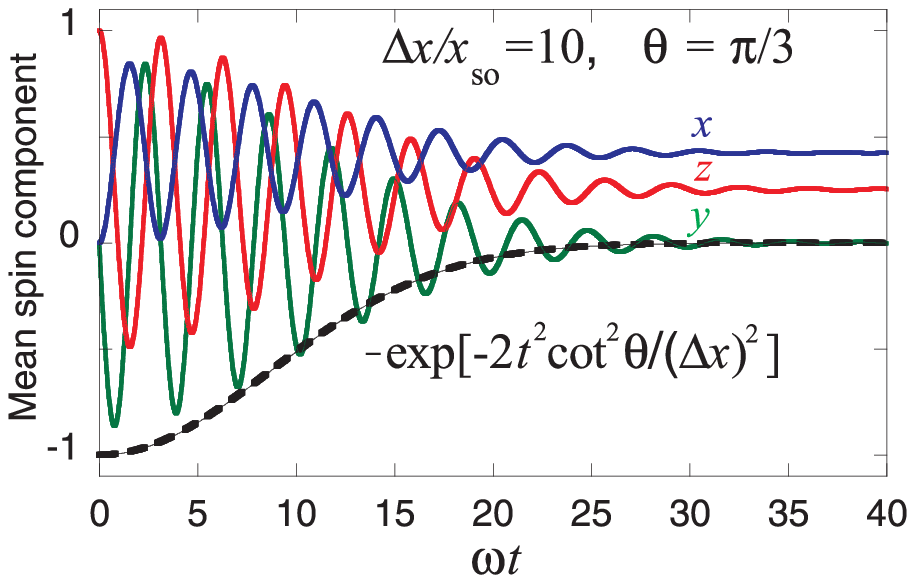}
\caption{(color online) Time dependence of $\overline \sigma_{x,y,z}$ (components are shown near the
lines) for a relatively weak spin-orbit coupling $\Delta x/x_{\rm so}=10$,
$\theta=\pi/3$ and $\et^{[{\rm in}]}=(1,0)^T$. The decoherence seen in $\overline\sigma_{y}$ 
is approximately Gaussian, as indicated by the dashed line. As expected 
from Eq.(\ref{12}) for a large $\Delta x/x_{\rm so}$,
we obtain $\lim_{t\to \infty}\overline {\sigma}_{z}(t)=1/4.$  Here we use physical units for length and time.}
\label{fig:FIG1}
\end{figure}

Taking the trace of the pure state
$|{\bm \Psi}(x,t)\ra\la{\bm \Psi}(x,t)|$ over the atom's position $x$, we find 
spin's reduced density matrix $\rho_s$  to be given by an incoherent  sum, 
\begin{equation}
\rho_s(t)=2 \pi  \int 
|A(p)|^2 | \u(p,t)\et^{[{\rm in}]}\ra \la\et^{[{\rm in}]} \u(p,t)|dp, 
\end{equation}
where each term corresponds to the
spin precessing in the composite magnetic field consisting of the original ${\mathbf B}$ plus a spin-orbit field induced by $p$ (cf. Ref. \cite{SG,SG1}). 
From (\ref{6}) we see that the matrix elements of $ \u(p,t)$ contain terms which oscillate as functions of $p$ ever more rapidly as the time increases.
For a smooth wave packet, such as a Gaussian one,
\begin{eqnarray}\label{11}
A(p)=\frac{(\Delta x)^{1/2}}{(2\pi)^{3/4}} \exp(-p^2(\Delta x)^2/4),
\end{eqnarray}
the integrals involving these oscillatory terms 
 will vanish as
$t\to \infty$, leaving the spin in a steady state $\rho_s(\infty)$ which, in general, 
depends on $\Delta x$ and $\et^{[{\rm in}]}$. 
The same applies to the averages of spin components $\bar \sigma_i(t) \equiv {\rm tr}[\sigma_i \rho_s(t)]$, which will reach steady values at long times. 
For example, 
for a spin initially polarized along the $z$-axis, 
\begin{eqnarray}\label{IN}
\et^{[{\rm in}]}=(1,0)^T,
\end{eqnarray}
 we have
\begin{eqnarray}\label{12}
&&{\lim_{t\to \infty}\overline {\sigma}_{z}(t)=\frac{\Delta x}{\sqrt{2\pi}}}\times \nonumber \\
&&\int_{-\infty}^{\infty}\frac{\exp(-p^2(\Delta x)^2/2)(p+\cos\theta)^2}{(p+\cos\theta)^2+\sin^2\theta}dp. 
\end{eqnarray}
The time dependence of $\overline \sigma_x$, $\overline \sigma_y$ and $\overline \sigma_z$ is shown in Fig.1.
The {type of decoherence in the spin subspace} depends on the direction of the external magnetic field, and varies from Gaussian, 
$\sim \exp(-{\rm const}\times t^2)$, for large $\Delta x$ and  $\theta\ne \pi/2$, 
to slow \cite{SG}, $\sim t^{-1/2}$ for narrow wave packets, or $\theta$ close to $\pi/2$.
{The reason for this behavior, partially illustrated in Fig.\ref{fig:FIG1}, can be seen as follows.
For example, the expression for the out-of-plane component $\overline{\sigma}_{y}(t)$ contains oscillatory integrals of the type
\begin{equation}
I_\pm =\int \frac{1}{\Omega}\exp\left[\pm2 i\Omega t -(\Delta x)^2p^2/2\right]dp, 
\end{equation}
with $\Omega$ defined in Eq.(\ref{6a}).
Expanding the phase $\Omega$ to the second order in $(p-p_s)$, $p_s=-\cos\theta$,  and evaluating the 
Gaussian integrals, yields 
\begin{eqnarray}\label{int}
&&\hspace{-1cm}|I_\pm|\sim \frac{\sin^{1/2}\theta}{\left[(\Delta x)^4\sin^2\theta+4t^2\right]^{1/4}} \times \\
&&\quad\exp\left[-\frac{2 t^2(\Delta x)^2\cos^2\theta}{(\Delta x)^4\sin^2 \theta+4t^2}\right]. \nonumber
\end{eqnarray}
Equation (\ref{int}) smoothly interpolates between the inverse power law $|I_{\pm}|\sim t^{-1/2}$
for $\theta =\pi/2$, and the Gaussian behavior  $|I_{\pm}|\sim \exp[-2t^2\cot^2 \theta/(\Delta x)^2]$
for  $\Delta x$ sufficiently large, at $\theta \ne \pi/2$ or $0$. In the trivial case $\theta=0$
all the fields are directed along the $z$-axis, there is no spin precession,
and $\sigma_z$ is conserved.
\newline
}

\section{Atomic density in the coordinate space} 

To study spatial distribution of the atomic cloud 
we begin by noting from Eq. (\ref{4}) that the mean anomalous velocity of the 
atom is given by $\overline {v}(t) = i\overline{[H,x]}=\overline{\sigma}_{z}(t)$, 
so that its mean position at a time $t$ is
\begin{eqnarray}\label{13}
\overline x(t) =\int_0^t \overline \sigma_z(t')dt'.
\end{eqnarray}
This is another quantum analog of the classical equation (\ref{3}). It shows that once the $z$-component of the spin 
settles into its stationary value (see Fig.\ref{fig:FIG1}),  the center of mass of the atomic cloud  $\overline x(t)$ will move with a constant velocity.

Equation (\ref{13}) does not, however,  tell us what the cloud looks like. To learn more about the cloud's shape we 
return to the convolution formula (\ref{7}).
It involves two known quantities: the wave packet $G(x-x',t)$ describing the atom {\it as it would be without the SO interaction}, and, via Eq.(\ref{10}),  
the amplitude distribution of $\la \sigma_z\ra_t$ for a spin precessing in the magnetic field ${\mathbf B}$, also {\it without the SO interaction}.
General properties of such distributions are known (see, e.g., \cite{ERG,DSD}), which simplifies our analysis.
In the following we will assume that the mass of the atom is sufficiently large to neglect 
the spreading of the wave packet (a justification will be given in Sect. VI below). Thus, we have 
\begin{equation}
G(x,t)\approx G(x,0)=\left[\frac{2}{\pi(\Delta x)^2}\right]^{1/4}\exp(-x^2/(\Delta x)^2).\q
\end{equation}
The advantage of applying Eq.(\ref{7}) is that we effectively 'look' at the details of the Green's function (\ref{8}) through a sort of a 
'microscope' with a resolution $\Delta x$. The smaller $\Delta x$ is, the more of the fine structure of $\et(x',t)$ is revealed. 
The larger is $\Delta x$ is, the less information about the distribution of $\la \sigma_z\ra_t$ we obtain.

It is convenient to introduce the  probability amplitudes 
to have $\la\sigma_z\ra_t=x/t$ provided the spin 
starts in the state $|j\ra$ and ends in the 
state $|i\ra$, 
\begin{eqnarray}\label{14}
u_{ij}(x,t)= (2\pi)^{-1}\int\exp(ipx)\u_{ij}(p,t)dp \\ \n
=(2\pi)^{-1}\exp[-ix\cos\theta]\xi_{ij}.
\end{eqnarray}
Deforming the integration paths in Eq.(\ref{8}) into contours in the complex $p$-plane, for $\xi_{ij}$ we find
 ($\beta\equiv \sin\theta$)
\begin{eqnarray}\label{15}
&&\xi_{11}(x,t)=\delta(x-t)+\chi(t+x)\chi(t-x)(1+x/t) \\
&&\qquad\times \int_{-\beta}^{\beta} \sin(t\sqrt{\beta^2-p^2})\exp(-px) dp = \xi_{22}(-x,t), \nonumber \\
&&\xi_{12}(x,t)=i\beta\chi(t+x)\chi(t-x) \nonumber \\
&&\qquad\times \int_{-\beta}^{\beta}\frac{ \cos(t\sqrt{\beta^2-p^2})}{\sqrt{\beta^2-p^2}}\exp(-px)dp =\xi_{21}(x,t),\nonumber
\end{eqnarray}
and $\chi(x)=1$ for $x\ge 0$ and $0$ otherwise. 

General properties of such amplitudes, as are mentioned above, are known in quantum measurement theory \cite{ERG,DSD}. 
The diagonal terms, $\zeta_{ii}$, have $\delta$-singularities responsible for the Zeno effect. 
The effect arises if, for a given $t$, the error with which one measures a time average, 
tends to zero. It locks the measured system in the eigenstates of the measured quantity,
in our case, $\sigma_z$.

For a { time that is long compared to the period of the Larmor precession, that is for $t\gg 1$,}
all $\zeta_{ij}(x,t)$ are highly oscillatory, with stationary regions near $x_{s}^{\pm}=t\la\pm \mathbf{n}|\sigma_z|\pm \mathbf{n}\ra$, where 
\begin{equation}
|\mathbf{n}\ra=\left[\cos\frac{\theta}{2}, \sin\frac{\theta}{2}\right]^T, 
|-\mathbf{n}\ra=\left[-\sin\frac{\theta}{2}, \cos\frac{\theta}{2}\right]^T, 
\end{equation}
are the spin states polarized along and against the {effective magnetic} field $\mathbf{B}$, respectively. The stationary points account for 
the ergodic property of quantum motion \cite{ERG}. If the accuracy $\Delta x$ is kept finite, while the measurement time increases, 
the unperturbed Hamiltonian $\h_0=\omega (\mathbf{n}\cdot \bm{\sigma})$ overcomes the restrictions imposed by the meter. 
The spin's density matrix becomes diagonal in the eigenbasis of $\h_0$, and the system explores its Hilbert space in such a way that 
the time average of a variable tends to its average in the instantaneous  von Neumann measurement made on this diagonal mixed state \cite{ERG}.

To prove the above, for $t\gg 1$
we evaluate the integrals in (\ref{15}) by the saddle point method, which yields 
\begin{eqnarray}\label{16}
&&\xi_{11}(x,t)=\delta(x-t)-\sqrt{1/2\pi \beta}(t+x)^{1/4}(t-x)^{-3/4} \nonumber \\
&&\qquad\qquad\times\cos[\beta(t^2-x^2)^{1/2}+\pi/4],  \nonumber \\
&&\xi_{12}(x,t)=i\sqrt{1/2\pi\beta}(t^2-x^2)^{-1/4} \nonumber \\
&&\qquad\qquad\times \sin[\beta(t^2-x^2)^{1/2}+\pi/4].
\end{eqnarray}
It is seen that $\zeta_{ij}(x,t)$ in Eq.(\ref{14}) have stationary points at 
\begin{eqnarray}\label{17}
x^{\pm}_{s}=\pm t\cos\theta.
\end{eqnarray}

\begin{figure}[h]
\includegraphics[scale=0.8]{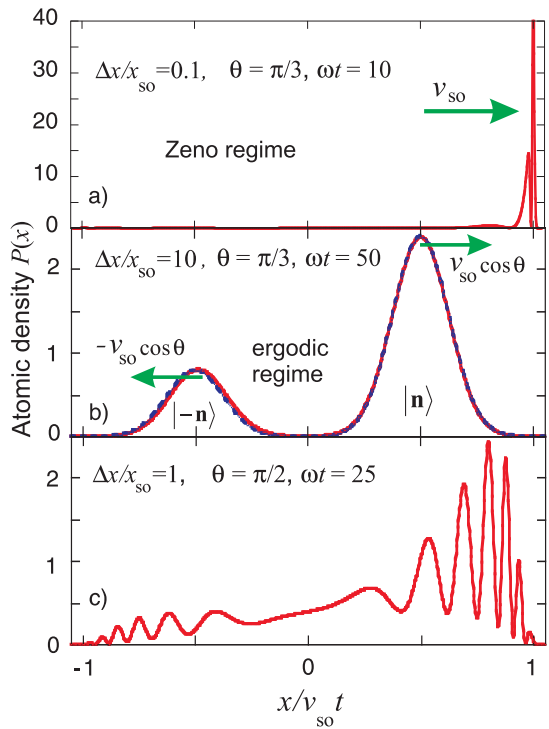}
\caption{(color online) 
Distribution of the atom's positions at a time $t$, $P(x)\equiv |\Psi_{1}^{2}(x,t)|+|\Psi_{2}^{2}(x,t)|,$ for $\et^{[{\rm in}]}=(1,0)^T$: 
a) in the Zeno regime the atom moves to the right at the speed $v_{\rm so}$;
b) in the ergodic regime the wave packet is split into two parts moving with the velocities $\pm v_{\rm so}\cos\theta$. 
The solid line is the exact numerical calculation, and the dashed line shows approximation (\ref{19}); 
c) interference fringes in the intermediate regime $\Delta x \sim x_{\rm so}$. 
Here we use physical units for length and time.}
\label{fig:FIG2}
\end{figure}

In our case, the accuracy of measuring $\la \sigma_z\ra_t$, $\Delta x/t$ changes with time, but 
so do the amplitude distributions $\zeta_{i,j}(x,t)$, and
one cannot say {\it apriori} whether the improvement in the accuracy is rapid enough to lock the system in the 
Zeno effect, or whether it would lead to the ergodic regime {involving stationary points of the propagator} 
described above. Next we will show that 
both situations are possible, depending on the ratio between the width of the wave packet $G(x,0)$, 
and the characteristic distance  $x_{\rm so}$. 
Thus, for a narrow initial wavepacket, $\Delta{x}\ll1$, and $t\gg 1$ from (\ref{12}) we have 
\begin{eqnarray}\label{18}
\overline {\sigma}_{z}(t)=1+O(\Delta x).
\end{eqnarray}
This is the Zeno effect - at long times the spin is locked in its initial state $|1\ra$ \cite{soc}. 
Accordingly, the time average  $\la \sigma_z\ra_t$ also tends to unity, with only the 
$\delta$-term in the first of Eqs.(\ref{15}) contributing to ${\bm \Psi}(x,t)$ in Eq.(\ref{7}).
As a result, the initial wave packet $G(x,0)$ travels as a whole to the right at a constant speed $v_{\rm so}$,
as can be seen in Fig.\ref{fig:FIG2}(a).

For $\Delta x \gg  1$ and $t\gg 1$ Eq.(\ref{12}) yields $\overline {\sigma}_{z}(t)\approx \cos^2\theta$.
The position of the atom is determined by the stationary points $x^\pm_s$ of the 
regular part of $\et(x,t)$. Expanding the phases of the cosine and sine in Eq.(\ref{16}) around $x_s^\pm$ to the 
second order in $x-x_s$ and evaluating resulting Gaussian integrals yields

\begin{eqnarray}\label{19}
{\bm \Psi}(x,t)\approx
C^+\exp[-\gamma^+(x-x^{+})^2]|\mathbf{n}\ra +
\\ \nonumber 
C^-\exp[-\gamma^-(x-x^{-})^2]|-\mathbf{n}\ra
\end{eqnarray}
where 
\begin{eqnarray}\label{20}
&&C^{\pm}=-\frac{\exp[\pm i (t+\pi/4)]\sqrt{\Delta x(1\pm \cos\theta)}}
{2^{3/4}\pi^{1/4}\sqrt{t^2\sin^2\theta \pm i(\Delta x)^2 t/2}}, \nonumber \\
&&\gamma^\pm=-\frac{(\Delta x)^2\pm 2it\sin^2 \theta}{4t^2\sin^4\theta+(\Delta x)^4}.
\end{eqnarray}
This is an example of the ergodic behavior - we have two Gaussians, each corresponding to the spin polarized along and 
against the magnetic field. The Gaussians move in the opposite directions with velocities proportional to 
the long time average $\la \sigma_z\ra_t$ for the spin in the $|\mathbf{n}\ra$ or 
$|-\mathbf{n}\ra$ state (see Fig. \ref{fig:FIG2}(b)). By ergodicity \cite{ERG}, these time averages equal the ensemble averages
$\la\mathbf{n}|\sigma_z|\mathbf{n}\ra$ and $\la-\mathbf{n}|\sigma_z|-\mathbf{n}\ra$. With $\Delta x \gg  \tan \theta$, 
the Gaussians do not overlap, and the spin's mixed stated becomes
\begin{equation}\label{21}
\rho_s(t\to \infty) \sim |\mathbf{n}\ra \cos^2\frac{\theta}{2}\la\mathbf{n}| + |-\mathbf{n}\ra \sin^2\frac{\theta}{2}\la-\mathbf{n}|, 
\end{equation}
so that $\overline {\sigma}_{z}=\cos^2(\theta/2)\la\mathbf{n}|\sigma_z|\mathbf{n}\ra+\sin^2(\theta/2)\la-\mathbf{n}|\sigma_z|-\mathbf{n}\ra$ 
equals $\cos^2\theta$, as is mentioned at the beginning of this paragraph. 

The atomic density in the intermediate regime $\Delta x \approx x_{\rm so}$ is shown in
Fig. \ref{fig:FIG2}(c). Here a measurement resolves the oscillations of 
$\xi_{ij}$ in (\ref{16}) with a period larger than $\Delta x$ 
as can be obtained directly from Eqs. (\ref{8}),(\ref{14}) and  (\ref{16}),
while finer details such as more rapid oscillations are lost in the convolution (\ref{7}). 

\section{Possible experiments}

An experimental realization {of the spin-orbit system described here} may consist in collecting a large number of spin-polarized non- or weakly 
interacting bosonic atoms in the ground state of a quasi one-dimensional trap with a harmonic 
confinement along the $x$-axis. 
 At $t=0$ the potential is collapsed, and the spin-orbit coupling
is switched on. Then, depending on the initial confinement and 
the direction of the external magnetic field one may be able to drive the condensate as 
a whole (see Fig. \ref{fig:FIG2}(a)), split it in two spin-polarized parts moving in the opposite directions 
(see Fig. \ref{fig:FIG2}(b)), or observe interference fringes similar to those shown in Fig. \ref{fig:FIG2}(c).

Now we return to physical units. To check the viability of the experiment, one may use as the energy scale the 
recoil energy $E_{r}=k_{r}^{2}/2M=2\pi^2/M\lambda^2$, where $\lambda$ is the 
wavelength of the laser light used to induce the spin-orbit coupling. 
For $^{87}{\rm Rb}$, one can achieve in experiment  $v_{\rm so}\sim k_{r}/M\sim 1$ cm/s \cite{Wang,Spielman,Engels}.
With the characteristic frequency of the trap, $\Omega_{\rm tr}$,  
of the order of $0.01E_{r}$ \cite{Wang,Spielman,Engels}, 
the width of the initial Gaussian state becomes $\Delta x=1/\sqrt{M\Omega_{\rm tr}}
\sim k_{r}^{-1}\sqrt{E_{r}/\Omega_{\rm tr}}\sim 10k_{r}^{-1}$. 
The spreading rate of the initial Gaussian, $v_{\rm wp}$, 
is then about $1/\Delta x M \sim \sqrt{\Omega_{\rm tr}/E_{r}}v_{\rm so}\sim 0.1 v_{\rm so}$.
As a result, the displacement of the wave packet due to spin-orbit coupling is much larger than the packet 
spreading, and the spreading does not influence the classification of asymptotic measurement regimes in Fig.\ref{fig:FIG2}. These can be
obtained for a constant wave packet width {since at given $\theta$ and $v_{\rm wp}\ll v_{\rm so}$ they are determined solely by the $\Delta x/x_{\rm so}$
ratio. Although the fringe pattern in Fig.\ref{fig:FIG2}(c) can be influenced by the packet
broadening at very long times, fringes will remain as a fingerprint of the intermediate regime.}
Taking into account that $x_{\rm so}=v_{\rm so}/\omega,$ we obtain
$\Delta x/x_{\rm so}\sim\sqrt{E_{r}/\Omega_{\rm tr}}\omega/E_{r}\gg\omega/E_{r}$. By changing
$\omega$ e.g., by tuning the phase of the laser fields, in the range from zero to $\lesssim E_r $, one can 
go from the Zeno to the ergodic regime. Finally, the repulsion 
between atoms can be neglected if the mean interaction energy per atom is 
small compared to $E_{r}$ - a condition easily achieved for typical Bose gases.

\section{a brief summary}

In summary, the mechanism behind manipulation of a cold atom cloud 
by switching on and off the spin-orbit interaction 
is that of a quantum measurement. 
The measurement, performed on the spin component 
coupled to the atom's momentum, can probe both the Zeno and the ergodic regimes.
Depending on the external Zeeman-like field, 
it can lead to different decoherence regimes for the monitored spin, 
ranging from a Gaussian decay to an algebraic law. The effects of the measurement on the atoms' density profiles 
include translation, spin-dependent fragmentation, and formation of interference fringes.

\section{Acknowledgements}  
We acknowledge support of the MINECO of Spain 
(grant FIS 2009-12773-C02-01), the Government of the Basque Country (grant
"Grupos Consolidados UPV/EHU del Gobierno Vasco" IT-472-10), and the UPV/EHU (program UFI 11/55).

\end{document}